\documentstyle[preprint,aps,epsf,floats]{revtex}
\begin{document}
\tighten

\preprint{\vbox{\hbox{JHU--TIPAC--97013}
\hbox{hep-ph/9707403}}}

\title{Summing $O(\beta_0^n \alpha_s^{n+1})$ Corrections to Top Quark Decays}
\author{Thomas~Mehen\footnote{address after Sept. 1 1997:
California Institute of Technology, Pasadena, CA 91125}}
\address{Department of Physics and Astronomy,
The Johns Hopkins University\\
3400 North Charles Street,
Baltimore, Maryland 21218 U.S.A.\\
{\tt mehen@dirac.pha.jhu.edu}}

\date{July 1997}

\maketitle
\begin{abstract}
Order $\beta_0^n \alpha_s^{n+1}$ QCD corrections to top quark decays into $W^+$ and $H^+$ bosons are computed to all orders in perturbation theory.  Predictions for the radiative corrections to the top quark width are compared with the estimates from BLM scale setting procedures.  The results of the summation are shown to greatly improve understanding of higher order corrections in the limit $m_W,~m_H \rightarrow m_t$, where the BLM scale setting method is known to fail. Attempts to reduce nonperturbative error by substituting the running mass for the pole mass in the expression for the decay are shown to fail in the limit $m_W,~m_H \rightarrow m_t$ because of subtleties in the treatment of phase space.

\end{abstract}

\pagebreak

Now that the top quark has been observed by CDF\cite{CDF} and D0\cite{D0}, future experiments will attempt to do precise measurements of its properties.  To date, only the leading order QCD corrections to the decays of the top quark into $W^+$ and $H^+$ bosons have been computed\cite{WDecay,HDecay}. The theoretical predictions suffer from considerable uncertainty since the renormalization scale for the strong coupling is undetermined at leading order. Corrections of $O(\beta_0 \alpha_s^2)$ are known\cite{AC2,Mehen} and can be used to set the scale in the leading order calculation via the scale setting method of Brodsky, Lepage and MacKenzie(BLM)\cite{BLM}. ($\beta_0$ is the first coefficient of the QCD beta function: $\beta_0 = (11 - 2/3~n_f)/4\pi$.) 
In a previous letter \cite{Mehen}, I showed that the BLM scale setting method,  when applied to the decays $t\rightarrow W^+ + b$ and $t\rightarrow H^+ + b$, gives unreasonably small scales in the limit $m_W,~m_H \rightarrow m_t$. Here I will perform a summation of $O(\beta_0^n \alpha_s^{n+1})$ terms in the perturbation series and find that the summation gives a much more reasonable picture of the effect of higher order corrections when the BLM scale setting method fails. For the case of $t\rightarrow W^+ + b$, $m_W \ll m_t$, so the summation of $O(\beta_0^n \alpha_s^{n+1})$ terms affords little improvement over BLM scale setting. However, the all order summation of $O(\beta_0^n \alpha_s^{n+1})$ terms could be phenomenologically relevant if it turns out that the top quark can decay into a new heavy particle such as a charged Higgs.  More generally, the BLM scale setting method is widely used as a tool for estimating the impact of higher order corrections on perturbative QCD predictions. For this reason, it is important to understand when it can fail and what means are available for improving estimation of higher order corrections when it does fail.

In the BLM scheme, the scale is chosen so as to absorb the $O(\beta_0 \alpha_s^2)$ term into the leading order correction:
\begin{eqnarray}\label{tdecay}
\Gamma(t\rightarrow W^+/H^+ + b) &=&\Gamma_0\left(1 + {2 \alpha_s(m_t) \over 3 \pi}\left(r_0  + (r_1 \beta_0 + \delta) \alpha_s + ...\right)\right) \nonumber \\
&=& \Gamma_0\left(1 + {2 \alpha_s(\mu_{BLM}) \over 3 \pi} (r_0   + \delta \alpha_s + ...)\right) 
\end{eqnarray}
where 
\[\mu_{BLM} = m_t~{\rm exp}\left( {-r_1 \over 2 r_0} \right), \] 
\[\Gamma_0(t\rightarrow W^+ + b) = |V_{tb}|^2 {G_F~m_t^3 \over 8\sqrt{2}\pi} \left(1 + 2 {m_W^2 \over m_t^2}\right) \left(1 -  {m_W^2 \over m_t^2}\right)^2, \]
\[\Gamma_0(t\rightarrow H^+ + b) = |V_{tb}|^2 {G_F~m_t^3 \over 8\sqrt{2}\pi} {\rm cot}^2 \beta \left(1 -  {m_H^2 \over m_t^2}\right)^2 .\]
(In these expressions, as well as throughout this paper, I will neglect the mass of the b quark.)  The resulting BLM scale turns out to be extremely small, especially as $m_W$ and $m_H$ approach $m_t$. For example, in the case of a 175 GeV top quark decaying into a 150 GeV Higgs and massless b quark, the computed BLM scale is 200 MeV.  Taken at face value, the BLM scale tells us that the radiative corrections to this process are uncontrolled even though the energy released in the decay, 25 GeV, is much greater than the QCD scale.  It is not clear that the extremely small BLM scales reflect a genuine physical effect, since the very large scales that result from BLM scale fixing appear to be a consequence of a small leading coefficient in the perturbative series. 

Since the scale dependence of the QCD prediction is an artifact of the finite order truncation of the perturbative series, any scale fixing scheme is essentially a guess of the size of higher order coefficients in the perturbative expansion. Inserting the one-loop expression for the QCD coupling constant,
\[ \alpha_s(\mu_{BLM}) = { \alpha_s(m_t) \over 1 + \beta_0 \alpha_s(m_t){\rm ln}(\mu_{BLM}^2/m_t^2)} ,\]
into Eq.~(\ref{tdecay}) and expanding in powers of $O(\beta_0^n \alpha_s(m_t)^{n+1})$, one finds that the series after BLM scale setting is equivalent to:
\begin{eqnarray}
\Gamma &= &\Gamma_0 \left(1 + {2 \alpha_s(m_t) \over 3 \pi} r_0 \sum_{n=0}^\infty \left( r_1 \over r_0 \right)^n \beta_0^n \alpha_s(m_t)^{n+1} +... \right) \nonumber \\
&= &\Gamma_0\left(1 + {2 \alpha_s(m_t) \over 3 \pi}{r_0 \over 1 - \beta_0 \alpha_s(m_t) (r_1/r_0)} + ...\right)
\end{eqnarray}
The BLM scale setting scheme can be thought of as a crude summation of $O(\beta_0^n \alpha_s^{n+1})$ terms in the perturbative series with a simple ansatz for the coefficients of these terms. This scale setting scheme suffers from obvious deficiencies. First of all, the size of these coefficients can be grossly overestimated (or underestimated) if the leading (or next-to-leading) coefficient is anomalously small. This is what appears to be going on in the case of top decay. In Ref.~\cite{Mehen} it is observed that the rapid decrease of the BLM scale as $m_W,m_H \rightarrow m_t$ appears to be a consequence of the leading order term becoming small rather than the next-to-leading term becoming large.  It is not obvious that the small leading coefficient implies the kind of growth of higher coefficients predicted by the BLM scale setting scheme.  Computation of higher order $O(\beta_0^n \alpha_s^{n+1})$ terms allows one to assess the perturbative series as a whole so as to be less sensitive to the vagaries of a single coefficient.  The second deficiency concerns the asymptotic behavior of the series. The $O(\beta_0^n \alpha_s^{n+1})$ term of the perturbative series contains a piece proportional to $n_f^n \alpha_s^{n+1}$ which comes from the insertion of $n$ fermion bubbles in the gluon propagator of the leading order correction. Direct computation of these graphs shows that their coefficients exhibit factorial growth at large orders, whereas the BLM scale setting scheme predicts that the coefficients of higher order terms in the series grow geometrically.  This geometric series diverges if $r_1/r_0 = 1/\beta_0 \alpha_s$.  This divergence is an artifact of the BLM scale setting method and the real $O(\beta_0^n \alpha_s^{n+1})$ series does not exhibit any such divergence.  Of course, the true $O(\beta_0^n \alpha_s^{n+1})$ series is asymptotic and so one must be careful to define what is meant by its sum.  This is done by means of a Borel transform, as will be discussed below. The result of this summation contains no spurious divergences.

It is important to acknowledge some of the shortcomings of the summation of $O(\beta_0^n \alpha_s^{n+1})$ terms in the perturbative series.  There is no physical limit of QCD in which these terms dominate. Nevertheless, one may expect these terms to make up a large part of the $O(\alpha_s^{n+1})$ correction because of the factor $(11-{2\over 3} n_f)^n$. Indeed, many next-to-leading order corrections in perturbative QCD calculations are dominated by the $O(\beta_0 \alpha_s^2)$ term, and so the computation of this piece gives a decent approximation for the complete $O(\alpha_s^2)$ correction. (For examples of this, see Refs.\cite{BLM,Examples}.)  At higher orders, the coefficients of the $O(\beta_0^n \alpha_s^{n+1})$ terms grow like $2^n n!$. The numerical enhancement of these terms suggests they may be a useful guide for estimating the size of higher order corrections. 

The decay rates for $t \rightarrow H^+ + b$ and $ t \rightarrow W^+ + b$ can written in the form:
\begin{equation}\label{DecayRate}
\Gamma(t\rightarrow W^+/H^+ + b) = \Gamma_0 \left( 1 + {2 \alpha_s(m_t)\over 3 \pi} \sum_{n=0}^\infty \alpha_s(m_t)^n(r_n\beta_0^n + \delta_n) \right)
\end{equation}
The $\delta_n$ are $O(\beta_0^{n-1} \alpha_s^{n+1})$ or smaller and will not be computed. 
The series  $\sum_{n=0}^\infty  r_n(\beta_0 \alpha_s(m_t))^n$ does not converge because $r_n \sim 2^n n !$ for large $n$.  The sum of the infinite series is defined by its integral representation in terms of its Borel transform:
\begin{equation}\label{Transform}
S_{\infty} \equiv {1 \over \beta_0 \alpha_s}\int_0^\infty du e^{-u/\beta_0 \alpha_s} B(u);
\end{equation}
\[ B(u) \equiv \sum_{n=0}^\infty {r_n \over n!} \beta_0^{-n} u^n \]
The integral representation for $S_\infty$ is still not well defined because the function $B(u)$ has poles on the positive real $u$ axis. These are the infrared renormalons. In order to define the integral, the contour must be deformed into the complex plane. The result of the integral has an (imaginary) ambiguity depending on whether one chooses to deform the contour above or below the real axis. If the Borel transform of the perturbative series has a pole at $u = u_0$, then the ambiguity in the integral representation for the series, which is the difference between going above or below the real axis, is just the residue of the pole at $u_0$:
\begin{equation}\label{error}
{\rm Im}~S_\infty \sim e^{-u_0/\beta_0 \alpha_s(m_t)} \sim \left( {\Lambda_{QCD} \over m_t} \right)^{2u_0}.
\end{equation}
The ambiguity in the sum of the series has the same size as a power correction one would expect to receive from non-perturbative QCD. It is customary to define the sum of the perturbative series as the principal value of the Borel integral and use the imaginary part of the Borel integral as an estimate of the uncertainty in the perturbative theoretical prediction. 
 
It is important to understand that the error estimated this way is usually an
{\it underestimate} of the true size of nonperturbative effects. One way to see why this is the case is to note that $\Lambda_{QCD}$ appearing Eq.~(\ref{error}) is the parameter in the one-loop expression for the running coupling, which is $\approx 100~{\rm MeV}$, while the intrinsic scale of nonperturbative QCD effects is typically closer to $1~{\rm GeV}$.  For observables with an operator product expansion, the size of the uncertainty in the perturbative series as defined by the imaginary part of the Borel transform usually underestimates the size of the lowest dimension hadronic matrix element which makes up the leading non-perturbative correction \cite{Sub}.

For processes in which there is no gluon self coupling at leading order, the Borel transform can be compactly expressed in terms of the one loop correction computed with finite gluon mass-squared, $\lambda^2$ \cite{BBB}:
\begin{equation}\label{Borel}
B(u) = -{{\rm sin}(\pi u) \over \pi u}\int_0^\infty d\lambda^2\left( {\lambda^2 \over m_t^2}e^{\rm C} \right)^{-u} r_0^{\prime}(\lambda^2),
\end{equation}
where $r_0^{\prime}(\lambda^2) = d r_0(\lambda^2)/d\lambda^2$. In this expression, C is a renormalization scheme dependent constant which turns out to be $-5/3$ in the $\overline{MS}$ scheme. The coefficients $r_n$ can be computed by taking derivatves of the Borel transform:
\begin{equation}\label{Coeff}
r_n = \beta_0^n {d^n \over du^n} B(u)|_{u=0}
\end{equation}
Inserting the expression for the Borel transform in Eq.~(\ref{Borel}) into Eq.~(\ref{Transform}) results in the following expression for $S_\infty$ \cite{BBB}:
\begin{equation}\label{ReSum}
S_\infty = -\int_0^{\infty} d \lambda^2 \Phi(\lambda^2) r_0^{\prime}(\lambda^2) + {r_0(\lambda_L^2) - r_0(0) \over \beta_0 \alpha_s}.
\end{equation}
In this expression,
\[ \Phi(\lambda^2) = {1 \over \beta_0 \alpha_s} \left( {1 \over \pi} {\rm tan}^{-1}\left[{1+\beta_0 \alpha_s {\rm ln}(\lambda^2 e^C/m_t^2) \over \beta_0 \alpha_s \pi}\right] - {1 \over 2}\right),\]
and 
\[ \lambda_L^2 = - m_t^2~{\rm exp}\left[-{1\over \beta_0 \alpha_s(m_t)} - C\right] = e^{-C} \Lambda_{QCD}^2.\]
The imaginary piece of the Borel transform comes from the term proportional to $r_0(\lambda_L^2)$. The uncertainty in the summation of the series is defined to be: 
\begin{equation}\label{UnCert}
\delta S_\infty = {{\rm Im} r_0(\lambda_L^2) \over \pi \beta_0 \alpha_s}
\end{equation}
For the decays $t \rightarrow W^+ + b$ and $ t \rightarrow H^+ + b$, the principal value of the Borel integral is approximately the same as the partial sum of the series truncated at its smallest term, and the error given by Eq.~(\ref{UnCert}) has nearly the same numerical value as the smallest term in the series. 

\begin{figure}
\epsfxsize=12cm
\hfil\epsfbox{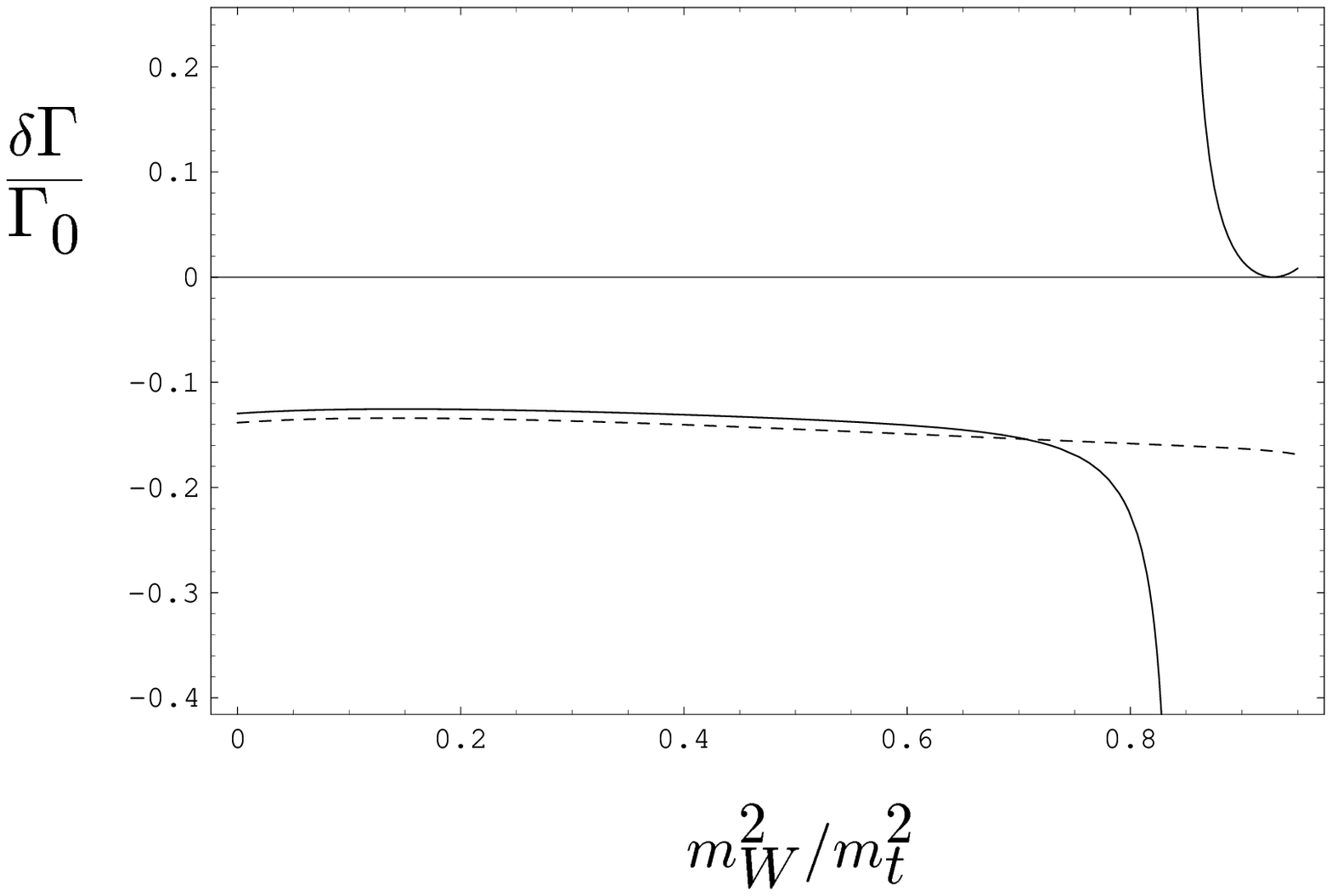}\hfill 
\caption{$\delta \Gamma/\Gamma_0$ for $t \rightarrow W^+ b$. Solid line - results after BLM scale setting; Dashed line - summation of $O(\beta_0^n \alpha_s^{n+1})$ terms} 
\label{delGam}
\end{figure}

In Fig.~\ref{delGam}, the result of applying the BLM scale setting method to $t \rightarrow W^+ + b$ is compared with the all orders summation of $O(\beta_0^n \alpha_s^{n+1})$ terms. I define $\Gamma = \Gamma_0 + \delta \Gamma$, and plot $\delta \Gamma/\Gamma_0$ as a function of $m_W^2/m_t^2$.  After applying the BLM procedure, $\delta \Gamma/\Gamma_0$ is given by:
\begin{equation}\label{BLMDG}
{\delta \Gamma \over \Gamma_0} = {2 \alpha_s(\mu_{BLM}) \over 3 \pi} r_0 =
{2 \alpha_s(m_t) \over 3 \pi} {r_0 \over 1 - \beta_0 \alpha_s(m_t) (r_1/r_0)}. 
\end{equation}
After summing all $O(\beta_0^n \alpha_s^{n+1})$ terms in the perturbation series, $\delta \Gamma/\Gamma_0$ is instead:
\begin{equation}\label{ReSumDG}
{\delta \Gamma \over \Gamma_0} = {2 \alpha_s(m_t) \over 3 \pi} S_\infty.
\end{equation}
The value of $\beta_0 \alpha_s(m_t)$ is taken to be 0.066.  For $m_W^2/m_t^2 \leq 0.7$, the BLM scale setting method gives a result that agrees fairly well with the all orders summation of $O(\beta_0^n \alpha_s^{n+1})$ terms.  For larger values of $m_W$, the BLM scale setting method ceases to give meaningful results because Eq.~(\ref{BLMDG}) has a pole near $m_W^2/m_t^2 = 0.85$. As discussed earlier, this spurious pole is an artifact of the geometric approximation of the full $O(\beta_0^n \alpha_s^{n+1})$ series.  When the higher order terms in the series are computed correctly the sum of the series exhibits no pole.  Perturbative corrections never exceed 17\% for all values of $m_W$.  

\begin{figure}
\epsfxsize=12cm
\hfil\epsfbox{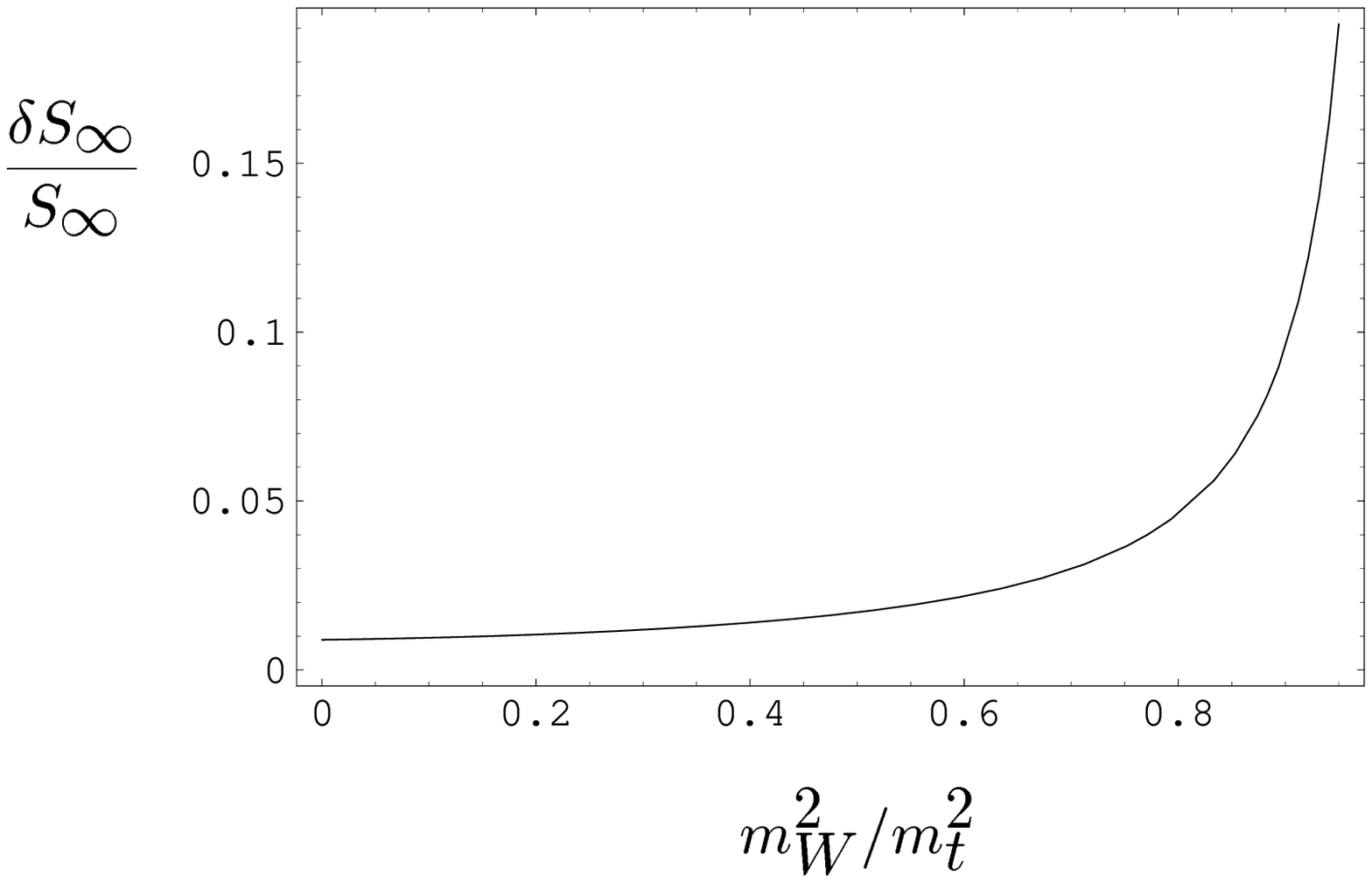}\hfill 
\caption{$\delta S_\infty/S_\infty$ for $t \rightarrow W^+ b$.} 
\label{delS/S}
\end{figure}

The uncertainty of the summation due to the asymptotic nature of the series is studied in Fig.~\ref{delS/S}, where I plot $\delta S_\infty/S_\infty$. As expected, the uncertainty increases as $m_W \rightarrow m_t$, but the uncertainty in the calculation of $S_\infty$ is never greater than 20\%.  Since this enters the decay rate multiplied  by a factor of $2 \alpha_s(m_t) /3 \pi = 0.023$, this amounts to an uncertainty in the prediction for the total rate that is less than $0.1 \%$.  As mentioned earlier, true nonperturbative corrections are expected to be larger. Nevertheless, it is safe to conclude that perturbation theory works well for top quark decay all the way to the kinematic limit. This conclusion is in stark contrast to what I find based on the application of the BLM scale. For $m_W^2/m_t^2 = 0.8, \mu_{BLM} = 484~{\rm MeV}$. One could be fooled in to thinking that perturbative QCD does not work. Examination of higher order $O(\beta_0^n \alpha_s^{n+1})$ terms indicates that it does.

For the physically relevant case of $m_W^2/m_t^2 = 0.21$, I find 
\[{\delta \Gamma \over \Gamma_0} = -0.135 \pm 0.006.\]  The estimate for the error is $\Lambda_{QCD}/m_t$ with $\Lambda_{QCD} = 1~{\rm GeV}$. This gives a significantly higher error estimate than what results if $\delta S_\infty$ is used. The reasons for this more conservative error estimate were given earlier.

\begin{figure}
\epsfxsize=12cm
\hfil\epsfbox{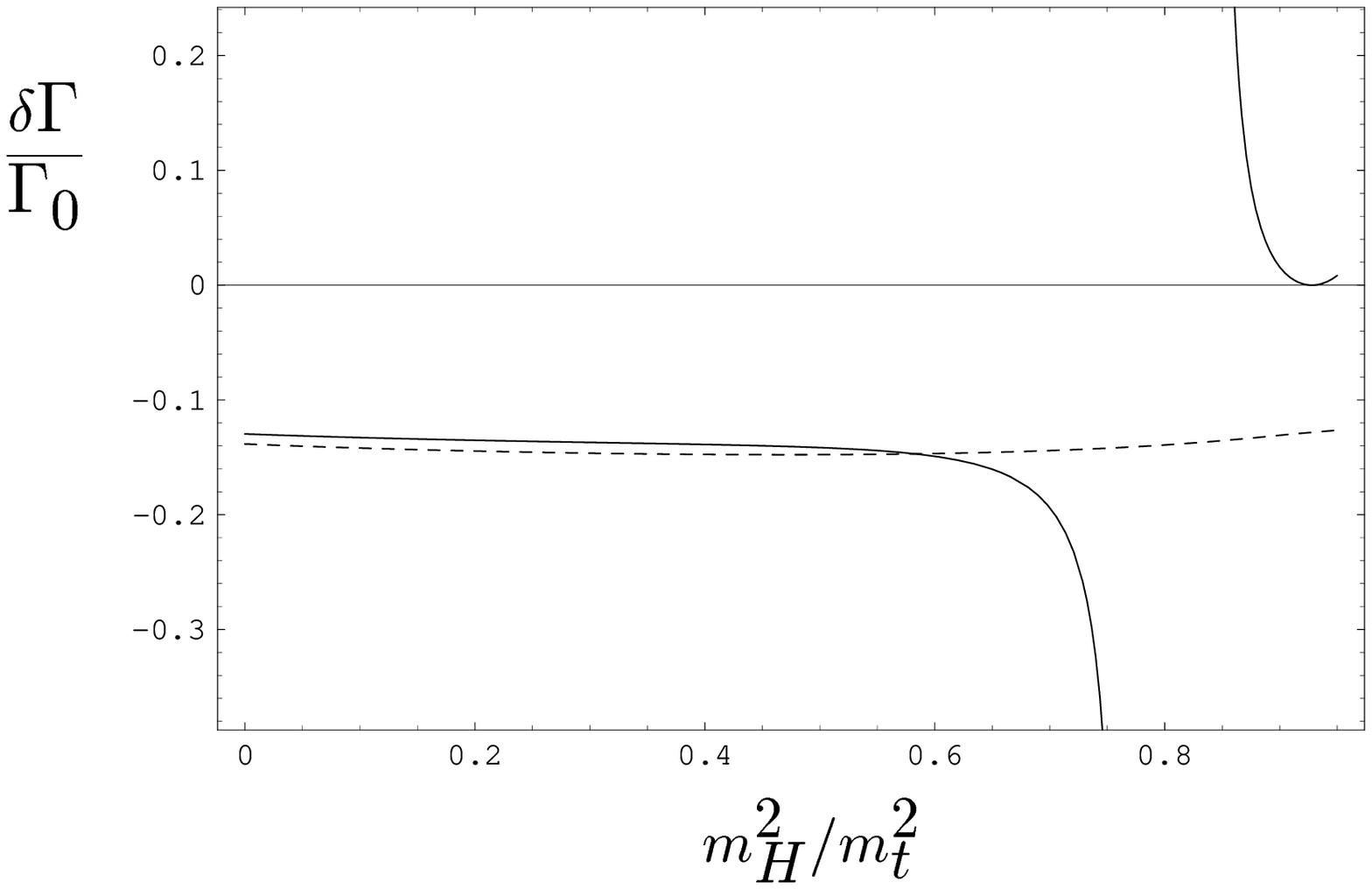}\hfill 
\caption{$\delta \Gamma/\Gamma_0$ for $t \rightarrow H^+ b$. Solid line - results after BLM scale setting; Dashed line - summation of $O(\beta_0^n \alpha_s^{n+1})$ terms} 
\label{dGHiggs}
\end{figure}

\begin{figure}
\epsfxsize=12cm
\hfil\epsfbox{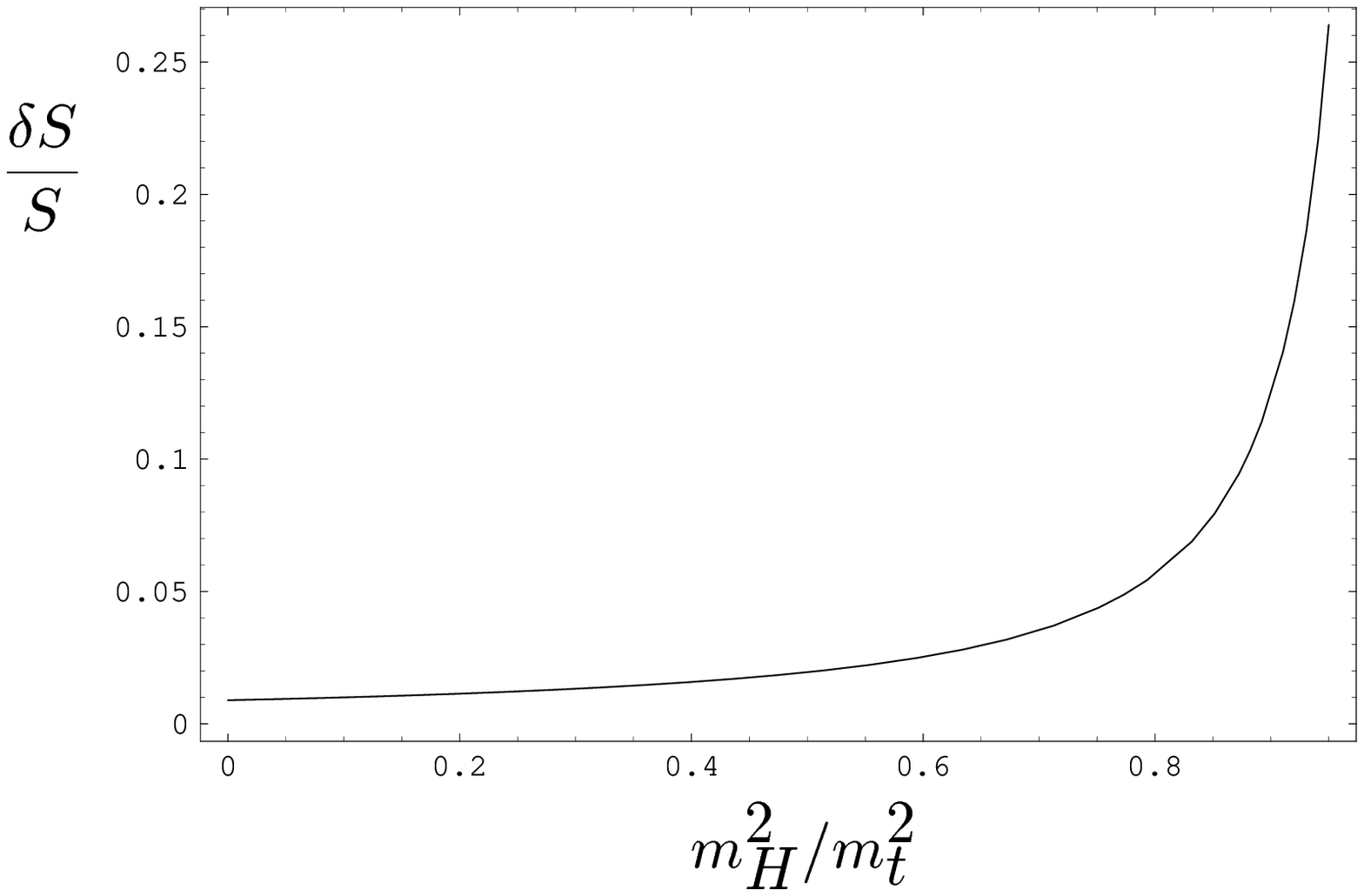}\hfill 
\caption{$\delta S_\infty/S_\infty$ for $t \rightarrow H^+ b$.} 
\label{dSSHiggs}
\end{figure}

The results for the decay $t \rightarrow H^+ + b$ are similar to the results for $t \rightarrow W^+ + b$ decay. In Fig.~\ref{dGHiggs} and Fig.~\ref{dSSHiggs}, I plot $\delta \Gamma/\Gamma_0$ and $\delta S_\infty/S_\infty$.  The BLM scale setting method breaks down for $m_H^2/m_t^2 \geq 0.6$, as opposed to $m_W^2/m_t^2 \geq 0.7$ in the case of W decay. The total correction to the decay rate is roughly 15\% for all values of the Higgs mass.  Uncertainties associated with the summation are approximately the same size as in the case of $t\rightarrow W^+ + b$, always less than 0.1\% of the total decay rate. Again, perturbation theory works for all values of $m_H$, despite the failure of the BLM procedure for large $m_H$.

The piece of the next-to-leading order corrections that has not been computed is expected to be $O(\alpha_s(m_t)^2/\pi^2) \approx 0.001$. This is actually smaller than the leading nonperturbative corrections of $O(\Lambda_{QCD}/m_t)$.  Unless the uncertainty due to nonperturbative corrections can be eliminated, computation of higher order corrections to top decay will afford no improvement of the theoretical prediction.  

In Ref.~\cite{Sub}, it is argued that rewriting the perturbative series in Eq.~(\ref{tdecay}) in terms of the top quark running mass will reduce the sensitivity of the perturbative series to long distance scales. The series relating the top quark pole mass to the top quark running mass (in the ${\overline{MS}}$ scheme) is given by
\begin{equation}\label{polemass}
m_t^{pole} = {\overline{m_t}}(m_t)\left(1 + {\alpha_s(m_t) \over 3 \pi} \sum_{n=0}^\infty d_n(\beta_0 \alpha_s(m_t))^n + ... \right),
\end{equation}
where $d_0 = 4, d_1 = 71/8 + \pi^2$, and higher order $d_i$ can be found in Ref.~\cite{BBB}. The Borel transform of this series suffers from a renormalon pole at $u = 1/2$, just like the series for top quark decay.  When the top quark pole mass is traded for the running quark mass, the poles at $u = 1/2$ cancel and the leading pole in the Borel plane is located at $u = 3/2$. Consequently, the error in summation of the series is $O(\Lambda_{QCD}^3)$. 

Replacing  $m_t^{pole}$ with the series invloving  ${\overline{m_t}(m_t)}$ results in the following expressions for top decay:
\begin{eqnarray}\label{RunMass}
\Gamma(t \rightarrow W^+ + b)&\sim&\overline{m_t}^3 \left(1 + 2 w\right) \left(1 -  w\right)^2  \left( 1+\sum_{n=0}^\infty (\beta_0 \alpha_s)^n \left(r_n^W + {3(1+ w + 2 w^2)\over 2(1+2 w)(1-w)}d_n\right)+...\right), \nonumber\\
\Gamma(t \rightarrow H^+ + b)&\sim&\overline{m_t}^3  \left(1 -  h\right)^2\left( 1+\sum_{n=0}^\infty (\beta_0 \alpha_s)^n \left(r_n^H + {(3+h)\over 2(1-h)}d_n\right)+...\right), 
\end{eqnarray}
where $w = m_W^2/\overline{m_t}^2$ and $h = m_H^2/\overline{m_t}^2$.
Note that in Eq.~(14) of Ref.~\cite{Mehen}, the expression for the series after trading the top quark pole mass for the top quark running mass is incorrect. Consequently, the calculations of the BLM scales for the series involving the running top quark mass in that paper are wrong. When the correct expressions in Eq.~(\ref{RunMass}) are used, the BLM scales are roughly $0.1-0.2~m_t$ and exhibit only a weak dependence on $m_W,~m_H$. 

\begin{figure}
\epsfxsize=12cm
\hfil\epsfbox{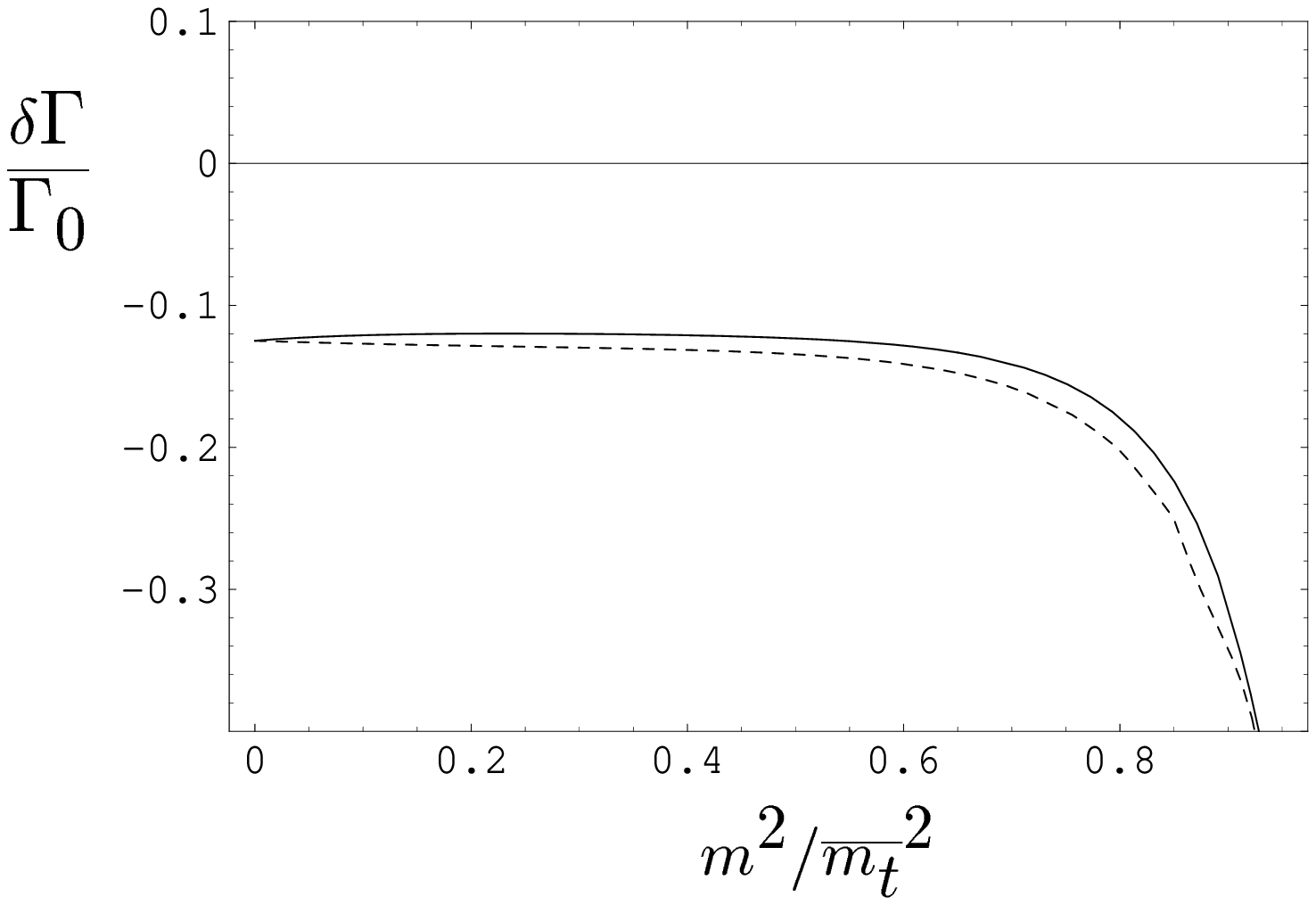}\hfill 
\caption{$\delta \Gamma/\Gamma_0$ after top quark pole mass is replaced with the running mass ($\overline{MS}$ scheme).  Solid line- $t \rightarrow W^+ + b$; Dashed line- $t\rightarrow H^+ + b$ line.}
\label{Running}
\end{figure}

After substituting the pole mass for the $\overline{MS}$ mass, I find the $\delta \Gamma/\Gamma_0$ shown in Fig.~\ref{Running}. In this case, the BLM scale setting method gives almost identical results as the all-order summation of $O(\beta_0^n \alpha_s^{n+1})$ terms.  The uncertainty in the summation $\delta S_\infty$ is negligible compared to uncomputed $O(\alpha_s^2)$ corrections for all values of $m_W,~m_H$.  The sum is dominated by the leading order coefficient in the perturbative series, the next-to-leading order term gives a small correction, and higher orders are negligble. This is why the BLM setting scheme gives such a good approximation to the summed series. 

For $m_W \ll m_t$, the prediction for the radiative correction is roughly the same whether one uses Eq.~(\ref{RunMass}) or Eq.~(\ref{DecayRate}). As $m_W \rightarrow \overline{m_t}$, the prediction for the radiative corrections becomes much larger if the series is expressed in terms of the running mass. 
This is because the phase space for the decay is incorrectly given in terms the running mass. The decay rate given by Eq.~(\ref{RunMass}) vanishes when $m_W,~m_H = \overline{m_t}$ instead of $m_W,~m_H = m^{pole}_t$. Trading the pole mass for the running mass results in an unphysical pole at $m_W,~m_H = \overline{m_t}$ in the expression for the radiative correction. Therefore, despite the formal reduction of non-perturbative errors, this substitution is not reasonable near the kinematic limit.

Even when $m_W \ll m_t$, it is questionable whether rewriting the decay rate in terms of the top quark running mass will really improve theoretical predictions for the top quark width. It is certainly true that the perturbative series for radiative corrections to the tree level rate is improved, in that terms higher order than $n=2$ in $O(\beta_0^n\alpha_s^{n+1})$ series are rendered negligible. The price to pay for this is that the tree level rate is now expressed in terms of a variable whose relation to physical observables suffers an $O(\Lambda_{QCD})$ ambiguity. The top quark mass experimentally defined as the peak in the invariant mass distribution of a W boson and b jet in a high energy hadronic collision is clearly the pole mass, up to corrections of $O(\Lambda_{QCD})$. If one tries to use Eq.~(\ref{polemass}) to determine the running quark mass from the experimentally determined top quark mass, the value of the running top quark mass will be subject to $O(\Lambda_{QCD})$ corrections, both from the fact the experimental top quark mass is not the pole mass, and because the series relating $m_t^{pole}$ and $m_t^{\overline{MS}}(m_t)$ suffers from an $O(\Lambda_{QCD})$ ambiguity.  Hence, the prediction for the top quark width will still recieve $O(\Lambda_{QCD})$ corrections.  Only if several observables are expressed in terms of the running mass, then the running mass eliminated to derive relations between these observables, will there be any hope of deriving theoretical predictions free from $O(\Lambda_{QCD})$ uncertainties.

In summary, I have discussed the improvement in the estimation of higher order corrections to the top quark decays $t \rightarrow W^+ +b$ and $t \rightarrow H^+ + b$. The BLM scale setting gives meaningless results as $m_W,~m_H \rightarrow m_t$, but summation of $O(\beta_0^n\alpha_s^{n+1})$ corrections shows that perturbation theory is well-behaved even in this limit.

I am indebted to Adam Falk for many enlightening discussions on the subject of perturbative QCD as applied to heavy quark decays. This work was supported by the National Science Foundation under Grant No. PHY-9404057.

\pagebreak

\begin{references}

\bibitem{CDF} F.~Abe {\it et al.}, Phys.\ Rev.\ {\bf D50} 2966(1994),
Phys.\ Rev.\ Lett.\ {\bf 74} 2626 (1995).

\bibitem{D0} S.~Abachi {\it et al.}, Phys.\ Rev.\ Lett.\ {\bf 74} 2632 (1995).

\bibitem{WDecay} M. Jezabek and J.H. Kuhn, Nucl.\ Phys.\ {\bf B314}, 1 (1989);
  Nucl.\ Phys.\ {\bf B320}, 20 (1989); Phys.\ Lett.\ {\bf B207}, 91 (1988).

C.S.~Li, R.~Oakes, and T.C.~Yuan, Phys.\ Rev.\ {\bf D43},3759 (1991).

A.~Czarnecki, Phys.\ Lett.\ {\bf B252} 467 (1990).

\bibitem{HDecay}A.~Czarnecki and S.~Davidson, Phys.\ Rev.\ {\bf D48} 4183 (1993);  {\bf D47} 3063 (1993).
 
C.S.~Li and T.C.~Yuan, Phys.\ Rev.\ {\bf D42} 3088 (1990); {\bf D47} 2156(E) (1993).

J.~Liu and Y.P. Yao, Phys.\ Rev.\ {\bf D46} 4183 (1992).

\bibitem{AC2} A.~Czarnecki, Acta Phys.~Pol.~B 26, 845 (1995) .

\bibitem{Mehen} T.~Mehen, Phys.\ Lett.\ {\bf B382} 267 (1996).

\bibitem{BLM} S.J.~Brodsky, G.P.~Lepage, and P.B.~Mackenzie, Phys.\ Rev.\ {\bf D28} 228 (1983).

\bibitem{Examples} M.~Luke, M.J.~Savage, and M.B.Wise, Phys.\ Lett.\ {\bf B345} 301 (1995); 

M.~Neubert, Phys.\ Rev.\  {\bf D51} 5924 (1995).
 
\bibitem{Sub} M.~Beneke, V.M.~Braun, and V.I.~Zakahrov, Phys.\ Rev.\ Lett.\ {\bf 73} 3058 (1994).

I.I.~Bigi, M.A.~Shifman, N.G.~Uralstev, and A.I.~Vainshtein, Phys.\ Rev.\ {\bf D50} 2234 (1994); 

\bibitem{BBB} M.~Beneke, V.M.~Braun, Phys.\ Lett.\ {\bf B348} 513 (1995).

P.~Ball, M.~Beneke, and V.M.~Braun, Phys.\ Rev.\ {\bf D52} 3929 (1995).


\end{references}
\end{document}